# Deadlock and Termination Detection using IMDS Formalism and Model Checking Version 2

Wiktor B. Daszczuk

Institute of Computer Science,
Warsaw University of Technology
wbd@ii.pw.edu.pl

**Abstract:** Modern model checking techniques concentrate on global properties of verified systems, because the methods base on global state space. Local features like partial deadlock or process termination are not easy to express and check. In the paper a description of distributed system in an Integrated Model of Distributed Systems (IMDS) combined with model checking is presented. IMDS expresses a dualism in distributed systems: server view and agent view. The formalism uses server states and messages. A progress in computations is defined in terms of actions consuming and producing states and messages. Distributed actions are totally independent and they do not depend on global state. Therefore, IMDS allows the designer to express local features of subsystems. In this model it is easy to describe various kinds of deadlock (including partial deadlock) and to differentiate deadlock from termination. The integration of IMDS with model checking is presented. Temporal formulas testing various kinds of deadlock (in communication or over resources) and termination are proposed.

## 1 Introduction

First attempts to deadlock detection were made on global state space of a centralized system [Haqu06, Zhou99] (or rather its model), by analysis of a graph of dependences called "wait-for graph". This approach allowed to predict a risk of deadlock statically. Alternate methods allowed to observe a system on-line in a selected "snapshots", which allowed to discover deadlocks in run-time [Chan85]. This is useful especially in systems in which global behavior cannot be predicted, like in a set of independent user programs requesting shared resources.

This approach was transferred to distributed systems with addition of locality obligation, as global state does not exist in general in such systems. Locality means that there are no notions like real time or simultaneity: a global decision of a deadlock state is made basing on independent local circumstances reported by system components [Chan83, Elma86, Mitc84]. The wait-for graph approach is successfully used till now, especially in run-time deadlock detection [Agar06, Knap87].

On the other hand, distributed termination detection techniques evolved [Huan89, Matt89, Peri04]. The methods are based on observation of special features of distributed processes (sometimes defined specially for termination detection) or control over message traffic.

Many modern system verification techniques are based on model checking, i.e. static exploration of global state space of a system (or a part of it in some techniques) [Clar99]. Many methods are used, typically based on temporal logics or similar formalisms. These techniques are exploded in research and lastly even in the verification of commercial software. Among the methods are graph-based (as statecharts [Hare87]) and language-based (as Promela [Holz95]). In verification, features are tested like safety (usually checked by "always" operator) and liveness (usually checked by "eventually").

The activities of the system are expressed in terms of local features of its components, and the global state space of the system is constructed. Many local features of system components may be expressed and verified by means of model checking. Deadlock is not a temporal property (it cannot be expressed by any temporal formula), but most model chceckers are equipped with deadlock detection procedures [Corb96, Puha00] , for example in SPIN [Holz95, Holz97, Have00]. Other approaches to identify deadlock by model checking are presented in [Kara91]. Usage of model checking techniques led to redefinition of deadlock, as it is difficult to express wait-for graphs in terms of temporal operators. Deadlock is usually defined as "state with no future", i.e. a strongly connected subgraph containing one state only: the deadlock itself [Kave01, Mage99].

This concerns total deadlock, when all components cannot make progress. Unfortunately, partial deadlock (in which there are processes than cannot continue, but there are still other processes that can run) can be neither expressed not found using model checking. Moreover, it is difficult to differentiate deadlock of a process from its termination (in both cases the process discontinues).

In the paper we will propose the application of Integrated Model of Distributed Systems (IMDS [Chro02]), which highlights locality of properties of verified system and exploits communication dualism (message passing/variable sharing). When combined with model checking, IMDS allows to express and find partial deadlock and to differentiate it from process termination. Moreover, the new method allows to distinguish a deadlock over resources from communication deadlock.

To sum up, "classical" model checking techniques do not allow to identify partial deadlocks and to differentiate deadlock from termination. The crisis comes probably from the fact that there is difficult to express partial deadlocks and partial terminations in terms of global state space. The feature that cannot be expressed, cannot be verified. Model checking combined with IMDS model allows to overcome these disadvantages.

Section 2 introduces informally a model of distributed system, in which deadlock and termination features may be easily expressed.  In section 3, an IMDS model for description of concurrent systems is described. Deadlock and termination will be expressed in terms of this formalism. Section 4 presents semantics of IMDS models (Labeled Transition System - LTS). Example verification of a system is described in section 5. Section 6 discusses various kinds of deadlock and termination.

## 2  The Model of Distributed System – An Overview

We will think of the distributed system as a set of *servers* communicating by means of *messages*. Every server has a set of variables. The values of these variables are called in whole a *server state*. Every server offers a set of *services* invoked by messages. An executed service (atomically) changes the state of the server.

The progress of calculations is defined by *actions*. An action is an execution of a service on a server in a context of a state of this server. A set of actions in a given calculation is called an *agent*. An action is invoked by a message in a context of an agent. The agent performs its calculation skipping between servers and calling their services. The agent is sequential – only one action is being executed at a time (or pending, if there are no conditions to execute an action). Nondeterminism is allowed.

An action on given server is defined for a given server state and a given agent calling a service on this server. The action once executed determines new server state. The server is sequential (at most one action is executed on a server at a time), but it may be nondeterministic as well (it may choose one of many possible actions to execute). The action typically produces a new message within the agent, to call a consecutive service on a determined server.

Additional assumptions are:
- a number of servers is fixed, every one of them having an initial state;
- given number of agents exist initially, every one of them issuing an initial message (calling a determined service on a determined server);
- an action may terminate an agent (then the next action for the agent is not determined).

Depending on the manner in which we connect the actions into sequences, we may express various aspects of distributed system:
1. We may observe a "server view" of the system, in which any process follows changes of an individual server states. A communication between server processes takes place by means of messages. Changes of states of the server are internal to its process, defining the progress of the server process.
2. We may observe an "agent view" of the system, in which any process follows the progress of an individual agents messages. Communication takes place by means of servers states. In this view messages cause a process to "skip" between servers, defining the progress of the agent process.

Other views are also possible, depending on the manner the actions are grouped [Chr06]. We will not deal with these additional views in this paper.

During the behavioral verification of distributed system, we often ask the questions on safety or liveness of the system. The questions may be asked in a temporal logic, and the model checker evaluates the truth of formulas. In our approach, we will ask formulas relative to the two views of a system:
- Agent safety is a feature that we expect to be always true in the agent view. A representative is agent safety from deadlock. As changes of servers states is the only way of agents' communication, we will call this kind of deadlock a *resource deadlock*. An agent falls into a deadlock if it waits for a server's state that cannot occur.
- Agent liveness is a property that – as we expect – to be reachable as long as the agent runs. Agent termination is a good example: the agent that is supposed to terminate does not fall into an endless loop and its termination is always reachable. In a case of never-terminating agent, its liveness should be defined as a property periodically and endless matched.

- Server safety is a feature that we expect to be always true in a server view. A representative is server safety from deadlock. As message passing is the only way of servers communication, we will call this kind of deadlock a *communication deadlock*. A server falls into a deadlock if it cannot change its state (forever) while there are unserved messages pending (unserved service calls). For example, if two servers wait for messages from each other, they both fall into a communication deadlock.
- Server liveness is a property that is expected to be always reachable (in server view). Server processes do not terminate for obvious reasons, therefore termination cannot be considered as server liveness feature. Instead, an invariant is a good example of liveness property (an invariant may be lost for a while, but it should be eventually restored every time). Of course there is not a "universal invariant" condition, it may be defined individually for any server adequately to its role in the whole system.

## 3   Integrated Model of Distributed Systems - IMDS

The process model IMDS is proposed by Chrobot and Daszczuk in [Chro06]. It contains elements that allow to observe server view or agent view of a distributed system in uniform framework.

Basic elements of a system are servers: (1)
- A system consists of named servers.
  *SVR* – set of system *servers*, $svr \in SVR$
- Any server has a state.
  *STAT* – set of *states*, $stat \in STAT$
- A server offers a set of named services.
  *SVC* – set of *services* to be called on servers, $svc \in SVC$

Agents run migrating between servers:
- Agent is a named calculation understood as set of service calls on servers (we call *a* an agent identifier).
  *AGT* – set of *agents* in the system, $a \in AGT$ (2)

A system may be described in terms of server states and pending messages (called services). Formally, a server state is a pair *(svr, stat)*:
  *SVST* = *{svr.stat | svr $\in$ SVR, stat $\in$ STAT}* (3)

A message is a triple *(svr, svc, a)*, "an agent *a* calls a service *svc* on a server *svr*":
  *MES* = *{a.svr.svc | a $\in$ AGT, svr $\in$ SVR, svc $\in$ SVC}* (4)

A Function *agent* retrieves agent component from a message, function *server* retrieves a server component from a message or a server state. Functions *agents*, and *servers* retrieve corresponding sets of agents and servers from sets of messages and server states.
  *a=agent(m), svr=server(m), svr=server(s), m $\in$ MES, s $\in$ SVR* (5)
  for *M$\subseteq$MES, agents(M)=$\cup_{m \in M}${agent(m)}*,
  for *M$\subseteq$MES, servers(M)=$\cup_{m \in M}${server(m)}*,
  for *S$\subseteq$SVST, servers(S)=$\cup_{s \in S}${server(s)}*,
  for *I$\subseteq$MES$\cup$SVST, servers(I)=servers(MES) $\cup$ servers(SVST)*.

(6)

For a given server *svr*, sets $MES_{svr}$ and $SVST_{svr}$ are respectively a set of all messages directed to it and set of all states residing on it, analogously $MES_a$ is a set of all messages with agent identifier *a*:

$$MES_{svr} = \{m \in MES \mid server(m)=svr\} \quad (7)$$
$$SVST_{svr} = \{s \in SVST \mid server(s)=svr\}$$
$$MES_a = \{m \in MES \mid agent(m)=a\}$$

A dynamic aspects of a system (progress) is defined in terms of actions. A system changes its state by means of execution of actions. An action has on its input: a state of a server and a message (which calls a service) in an agent. Actions are defined in terms of server states and messages. Every action has two input arguments: a server state and a message. The set *D* contains all pairs containing one server state and one message:

$$\text{For } M \subseteq MES, S \subseteq SVST, D(M,S)=\{\{m,s\} \mid m \in M, s \in S\} \quad (8)$$

The executed action consumes its input server state and its input message (they disappear from the system). Then, a new server state and (typically) a new message are caused (see later (10)).

The set *H*, representing the configurations of the whole system, contains the sets of states of all existing servers and messages in all existing agents. Therefore, in an element of *H*, no pair states may refer to the same server (any server *svr* has its state strictly defined any time) and no pair of messages may refer to the same agent (every agent *a* has its point in the calculation strictly defined as a single service *svc* called on a given server *svr*).

$$H(MES \cup SVST) \subseteq 2^{MES \cup SVST} \quad (9)$$
$$H(MES \cup SVST) = \{ X \subseteq MES \cup SVST \mid \forall_{m,m' \in MES} m \neq m' \Rightarrow agent(m) \neq agent(m'),$$
$$\forall_{s,s' \in SVST} s \neq s' \Rightarrow server(s) \neq server(s') \}$$

The action $\lambda$ is a relation between a pair from *D* and a configuration from *H*.

$$ACT \subseteq D(MES,SVST) \times H(MES \cup SVST); \quad (10)$$
$$ACT = \{\lambda \mid \lambda = <\{m,s\}, CI >, \{m,s\} \in D(MES,SVST), CI \in H(MES \cup SVST) \}$$

For any $\lambda = <\{m,s\}, CI > \in ACT$:
    a) $server(m) = server(s)$      // an action is executed on given server (11)
    b) $\exists_{cs \in (CI \cap SVST)} server(s)=server(cs), cs \in CI$
         // there is an obligatory continuation state (12)

Features a) and b) concern every action. A type of an action depends on an output set *CI*. An action is either regular or a terminating one: a regular action produces a new state of the server it is executed on (*continuation server state* $cs \in CI \cap SVST$, $server(cs)=server(s)$), and a message with the same agent identifier as the input message (*continuation message* $cm \in CI \cap MES$, $agent(cm)=agent(m)$). The continuation message represents a next step in the agent progress. Of course, a continuation state is exactly one (as no pair of states may exist on the same server in $CI \in H$ (from (12) and (9)) and a continuation message may be at most one (as no pair of messages may have the same agent identifier in $CI \in H$ (from (9)).

The terminating action produces no message with the same agent identifier as input message (no output message in the agent), $CI=\{cs\}$; note that output server state is obligatory (new state of the server).

A *configuration* $\gamma \in H$ is a set of messages and states in which every message is directed to a server present in this configuration (with its current state).

$\quad \gamma \in H(MES \cup SVST)$ is system configuration iff *servers($\gamma \cap MES$)* $\subseteq$ *servers($\gamma \cap SVST$)* 

$\hfill (15)$

This means that no message $m=a.svr.svc \in MES$ may be directed to the server *svr* whose state is not represented by any $s=svr.stat \in SVST$:

$\quad m \in \gamma \cap MES \Rightarrow \exists_{s \in \gamma \cap SVST}\ server(s)=server(m)$ (from (15)). $\hfill (16)$

Configuration $\gamma_0$ is an initial one. It contains initial states of all servers and "starting messages" of agents initially present. All servers states in a configuration $\gamma$ are *present* and all messages in $\gamma$ are *pending*.

When input state and input message of an action are present in $\gamma$, the action is *enabled* in $\gamma$ ($\lambda=<\{m,s\}, CI>$ is enabled in a configuration $\gamma$, iff $\{m,s\} \subseteq \gamma$). We say that an action $\lambda$ transforms a system configuration $\gamma$ into a new system configuration $\gamma'$ such that

$\quad \{m,s\} \subseteq \gamma$ and $CI \subseteq \gamma'$ and $\gamma - \{m,s\} = \gamma' - CI$ $\hfill (17)$

(all servers and agents not taking part in the action $\lambda$ preserve their states and messages[1]). Action execution is called *firing* the action. Many actions may be enabled on a server, but only one may be fired at a time (from (17)).

When a message $m \in \gamma \cap MES$ is pending on a server *svr* (*server(m)=svr*), the following situations may occur (from (10)-(16)):

- No action is defined on the server for the message *m*:
  $\forall_{\lambda \in ACT}\ \lambda=<\{m,s\}, CI>, s \in SVST, server(s) \neq svr$ $\hfill (18)$
  (this corresponds to a programming error – we will not deal with such cases).

- There is a state $s \in SVST$ paired with *m* (there exists an action $\lambda=<\{m,s\}, CI>$), and *s* is present in $\gamma$ ($\{m,s\} \subseteq \gamma$), then the action $\lambda$ may be executed (it is not inevitable because other messages $m',m'',... \in MES$, all with the same server identifier as *m*, concurring wit *m* to be fired).

- There are states $s,s',... \in SVST$ paired with *m*, and none of these states $s,s',...$ is present in $\gamma$,
  $\forall_{\lambda \in ACT}\ \lambda=<\{m,s\}, CI>, s \in SVST, server(s)=svr, s \notin \gamma$ $\hfill (18a)$
  but there is a sequence of actions[2] leading to a configuration $\gamma'$ in which some *s*, *server(s)=svr* is present; then the action $\lambda$ may be executed in future (once it is enabled). The agent *a=agent(m)* simply waits for other agents to change the state of the server *server(m)*, in which the action $\lambda$ would be enabled (and thus it may be fired).

- There are actions defined for the message (paired with some states:
  $\exists_{\lambda \in ACT}\ \lambda=<\{m,s\}, CI>, s \in SVST, server(s)=svr$ $\hfill (19)$
  ), but neither such state *s* is present in $\gamma$ nor it may occur in the future: no sequence of actions leads to a configuration $\gamma'$ in which *s* is present (this corresponds to a deadlock).

---

[1] This assumption leads to interleaving semantics in which at most one action is executed at a time. Other semantics with parallel actions is also possible
[2] Labeled transition system over IMDS will be defined in chapter 4

- There is present a paired state $s \in \gamma \cap SVST$, $\lambda = <\{m,s\}, CI>$ (the action $\lambda$ is enabled), but there are other enabled actions enabled in $\gamma$ (
  $$\exists_{\lambda' \in ACT} \lambda' = <\{m',s\}, CI'>, m' \in \gamma \cap MES, agent(m') \neq agent(m); \qquad (20)$$
  calling the same or other service on server *svr*); an action to execute is chosen in nondeterministic way from all such $\lambda, \lambda', \lambda'', \ldots$.

When a state $s \in \gamma$ is present on a server *svr* (*server(s)=svr*), the following situations may occur (from (10)-(16)):
- No message is pending on a server: for every $m \in \gamma$, *server(m)≠svr* – the server waits for a service to be called on it.
- There are pending messages on a server *svr*, but none of them is paired with the present server state *s* (
  $$\forall_{\lambda \in ACT} \lambda = <\{m,s'\}, CI>, m \in \gamma \cap MES, server(m)=svr, s' \notin \gamma \qquad (21)$$
  ) – the server waits (no action may be executed on a server at the time).
- There are messages paired with the present server state *s* (
  $$\exists_{\lambda \in ACT} \lambda = <\{m,s\}, CI>, m \in \gamma \cap MES, server(m)=svr, s \in \gamma \qquad (22)$$
  - there are actions enabled on the server *svr*) – one of such actions $\lambda, \lambda', \lambda'', \ldots$ is chosen to execution in nondeterministic way.

The assumptions made ((10)-(16)) lead to the following findings:
- In general an action has two preceding actions (except the first action on server/in agent): one delivers a server state $s \in SVST$ and second delivers a message $m \in MES$; in a particular case, a server state *s* and a message *m* may be delivered by a single preceding action
  $$\lambda_{pred}=<\{m',s'\}, CI_{pred}>, cm_{pred}=m \in CI_{pred} \cap MES, cs_{pred}=c \in CI_{pred} \cap SVST,$$
  $$server(cm_{pred})=server(cs_{pred}). \qquad (23)$$
- An action may have many consecutive actions possible: one on the same server (caused by continuation server state), one for every new server started in the "new server" action and one on every server called by output messages (continuation message and messages generated in "new agent" action).
- Typically a regular action has two consecutive actions if the output message is directed to other server:
  $$\lambda=<\{m,s\}, CI>, cm \in CI \cap MES, server(cm) \neq server(s). \qquad (24)$$
  Only one consecutive action is in the case in which *server(cm)=server(s)*.

The very important feature of actions is their *locality*. Actions on distinct servers are executed totally independently, as input states taking part in these actions have distinct server attributes. They neither concur nor inhibit each other. No system element external to a server influences which actions may be executed on it. This feature enables the designer to express desired features of a system in terms of local actions only. Moreover, proved features of a subsystem hold independently of the behavior of various wider-scale systems the subsystem works in (additional servers may be added to the specification and additional agents may be added, provided that they do not deliver server states occurring as input in actions of existing agents).

Now let us describe processes in the system.
- A *process* is a set of server states and messages.
- A *sequential process SP* is a set of states/messages connected by actions, for $\lambda=<\{m,s\}, CI>$, either *m* or *s* (or both) belongs to process *SP*, and either a continuation message or a continuation state belongs to the same process *SP*. For infinite process if $e \in SP$, $e \in \{m,s\}$:

$$\exists_{\lambda \in ACT}\ \lambda=<\{m,s\}, CI>,\ \exists_{i \in \{m,s\}}\ i \in SP\ \land\ \exists_{j \in CI}\ j \in SP \tag{25}$$

If no such action exists, the process terminates. Note that the process need not be deterministic: more than one action may be defined for the state/message $e$.

- *Synchronous sequential processes SSP* – one *custodian* process of a server *CSSP* delivers a state and second *messenger* process *MSSP* delivers a message:

$$\lambda=<\{p,s\}, CI>,\ p \in MSSP \cap MES,\ s \in CSSP \cap SVST. \tag{26}$$

The action takes place in both processes (custodian and messenger). Of course, for continuation message and state: $cm, cs \in CI$, $cm \in MSSP \cap MESS$ and $cs \in CSSP \cap SVST$. We will not consider such processes in the paper[3].

- *Asynchronous sequential processes ASSP* – a single process delivers both a state and a message to an action:

$$\lambda=<\{m,s\}, CI>,\ m,s \in ASSP. \tag{27}$$

Therefore, synchrony is hidden inside a process. The action takes place in the process delivering both the state and the message. The next action in the process is defined by the continuation message $cm \in CI \cap MES$, $cm \in ASSP$ or the continuation state $cs \in CI \cap SVST$, $cs \in ASSP$, depending on the type of process (described later).

- *Traveler T* – asynchronous sequential process which progress is determined by continuation message $cm$:

$$\lambda=<\{m,s\}, CI>,\ cm \in CI \cap MES,\ m,s,cm \in T \tag{28}$$

(it "travels" between servers as usually a continuation message has a service identifier distinct from an input message). A continuation message $cm$ determines a next action in the traveler $T$ on other server $svr$ (in a pair with some $s'$, $server(cm)=server(s')=svr$), or the traveler $T$ terminates. Travelers communicate by sharing the states of the servers (continuation state $cs \in CI \cap SVST$ is delivered as input state to an action in other traveler). Sometimes the continuation state $cs$ may be delivered to the traveler $T$ itself if an input message and a continuation message have the same server identifier: $server(m)=server(cm)$.

- *Resident R* – asynchronous sequential process which progress is determined by continuation state $cs$:

$$\lambda=<\{m,s\}, CI>,\ cs \in CI \cap SVST,\ m,s,cs \in R \tag{29}$$

(it "resides" on a server as the continuation state has the same server identifier as the input state, $server(cs)=server(s)=svr$). A continuation state $cs$ determines a next action in the resident $R$ on the server $svr$ (in a pair with some $m'$, $server(cs)=server(m')=svr$). Residents communicate by message passing (continuation message $cm \in CI \cap MES$ is delivered as input message of an action in other resident – running on other server $svr' \neq svr$). Sometimes the continuation message $cm$ may be delivered to the resident $R$ itself if an input message and a continuation message have the same server identifier: $cm \in CI \cap MES$, $server(cm)=server(m)$.

Every system described in terms of IMDS may be decomposed into:
- *travelers* (agent view) communicate by server states sharing; every traveler is built over messages with common agent identifier, and states of visited servers (or simply all servers in the system); it "travels" between servers,

---

[3] They are discussed in [Chro02]

- *residents* (server view) communicate by message passing; every resident is built over the states of a given server and all messages directed to this server; it "resides" on a server).

The two mentioned decompositions are canonic – the details and proof can be found in [Chro06]. As a continuation state is obligatory, therefore a resident process never terminates (a server exists forever). On the other hand, a continuation message may be absent (in special action "termination"). In such a case, an agent identifier is annihilated and the traveler terminates. From the definition of travelers and residents (28,29):

- For any $a \in AGT$, $T_a = MES_a \cup SVST_{servers(MES_a)}$           (30)
  is the *traveler process* with the agent identifier $a$ (if we include all server states into any traveler, $T_a = MES_a \cup SVST$).
- For any $svr \in SVR$, $R_{svr} = SVST_{svr} \cup MES_{svr}$           (31)
  is the *resident process* with the server identifier $ser$.

The traveler decomposition and resident decomposition are crucial to our target: identification of resource deadlock, communication deadlock and termination, as the two decompositions express the resource sharing and message passing aspects of verified system. The extraction of resource sharing/message passing activities for verification purpose is natural in IMDS. Attributes of travelers and residents are used directly in temporal formulas in model checking over IMDS. Such features are absent in other specification models and languages like Petri nets, Promela or Estelle. Of course, such languages may be used if they adopt notions of servers, states, services, agents and actions of IMDS.

## 4 Model Checking of IMDS

For model checking purpose, a system modeled in IMDS must be expressed as finite state Labeled Transition System. In LTS, system configurations represent its states (not to be confused with server states, therefore we call it system states) while actions play role of transitions.

$LTS = <\Gamma, \Lambda, \Theta>$           (32)

$\gamma_{max}$ – maximal configuration:

$\forall \gamma_{max}, \gamma_{max}' \in H(MES \cup SVST), \gamma_{max} \neq \gamma_{max}' \; \gamma_{max} \not\subset \gamma_{max}'$ and $\gamma_{max}' \not\subset \gamma_{max}$

$\Gamma$ – set of *system states*, $\Gamma = \{\gamma_{max} \in H(MES \cup SVST)\}$

$\Lambda$ – set of *system labels*, $\Lambda = \{\lambda = <\{m,s\}, CI>\}$

$\Theta$ – set of *system transitions*, $\Theta = \{\tau = <\gamma_{max}, \lambda, \gamma_{max}'> \mid \lambda = <\{p,s\}, CI>$,
$\{p,s\} \subseteq \gamma_{max}, \gamma_{max}' = (\gamma_{max} - \{p,s\}) \cup CI\}$

We must take additional assumption that after a number of new servers dynamically created (by special actions "new server"), no further server creation may take place. Also, after a number of agents created dynamically, a number of new agents created (by special actions "new agent") must be equal to a number to annihilated agents (agent identifiers should be reused after annihilation). These assumptions guarantee the finiteness of LTS.

In such an environment, some classes of deadlocks in distributed environment may be defined [Chro01] and easily differentiated from termination. For example, a traveler process $T_a$ is in deadlock over resources if a message with agent identifier $a$ is pending (a traveler process is not finished) and no action with this message on input can be enabled in future. The termination condition of $T_a$ differs: no message with agent identifier $a$ is pending (an agent disappears in an action). On the other hand, a resident process $R_{svr}$ is in a communication deadlock if no action can be enabled on the process' server *svr* while there are pending messages *m,m',m",...* directed to the server *svr* (*server(m)=server(m')=server(p")=svr*). To verify such properties, some model checking formalism may be applied in which the messages, server states and actions should be present or adopted (not to lose the carriers of definition of deadlock and termination).

Semantic issues of IMDS are following: enabled actions on distinct servers are always independent on each other (they use messages with different server identifiers and states of different servers). Enabled actions on a single server are always in conflict: they compete for a common state (a present one).

For model checking purpose, we should use a particular verification environment with its input language, and we must translate actions into this language. Although many environments are potentially available, we used two of them: COSMA environment, developed in ICS WUT [Dasz07] and Spin []. The translation of IMDS to CTL logic, used in COSMA, is described in [Dasz08]. Below is presented the translation to Promela, the Spin modeling language. Spin uses the LTL logic. The forms of CTL and LTL formulas checking deadlocks and terminations are very similar.

## 5   Example deadlock detection

Let us consider a concurrent system in which two agents use two semaphores (*m* and *n*, each one residing on its own server).

A1:  
*sem1.wait;*  
*sem2.wait;*  
*sem1.signal;*  
*sem2.signal;*  
*stop*

A2:  
*sem2.wait;*  
*sem1.wait;*  
*sem2.signal;*  
*sem1.signal;*  
*stop*

This system falls to a total deadlock while *A1* holds *sem1* and waits for *sem2* and *A2* holds *sem2* and waits for *sem1*. But the situation changes if we add another agent process *A3* which simply loops and does his own calculations. Processes *A1* and *A2* cannot continue, but *A3* still works. We denote calculations made by *A3* symbolically as endless execution of *left* and *right* services or the server *r*:

A3:  
*loop {*  
    *r.left;*  
    *r.right*  
*}*

The system is not in a total deadlock, it is a partial deadlock which are hard to identify using model checking techniques (no system state is a "state without future"). For the same reason, termination of *A1* and *A2* (which are designed to terminate after a number of calculations) is not inevitable (a deadlock is not a termination).

Let us define the example system in IMDS basic notation, server view, which is simply a grouping of actions on individual servers. This is a sequence of server specifications (enclosed by ***server** ... **end***), server and agent instances declaration (***agents*** …, ***servers*** …) and an initial phrase (***init*** → *[...]*). Each server has a set of states, a set of services and a set of actions assigned (in arbitrary order, every action specifies a consecutive action explicitly). In a definition of an action: *[fs, fr]* → *[res]* the header consists of a message, *fs*, a state, *fr*, and the result, *res*. The result is a set of (may be zero) messages and at least one state: continuation state. In an action, a message *fs* and a state *fr* are processed to create a set *res* of messages and server.

| | | | |
|---|---|---|---|
| **server:** | sem (**agents** A[2]; **servers** proc[2]), | | |
| **services** | {wait, signal}, | | |
| **states** | {up, down}, | | |
| **actions** | <j=1..2>{A[j].sem.wait, sem.up} | -> {A[j].proc[j].ok_wait, sem.down}, | |
| | <j=1..2>{A[j].sem.signal, sem.down} | -> {A[j].proc[j].ok_sig, sem.up}, | |
| | <j=1..2>{A[j].sem.signal, sem.up} | -> {A[j].proc[j].ok_sig, sem.up}, | |
| **end;** | | | |
| **server:** | proc (**agents** A; **servers** sem[2]), | | |
| **services** | {start, ok_wait, ok_sig}, | | |
| **states** | {ini, first, sec, stop}, | | |
| **actions** | {A.proc.start, proc.ini} | -> {A.sem[1].wait, proc.first}, | |
| | {A.proc.ok_wait, proc.first} | -> {A.sem[2].wait, proc.sec}, | |
| | {A.proc.ok_wait, proc.sec} | -> {A.sem[1].signal, proc.first}, | |
| | {A.proc.ok_sig, proc.first} | -> {A.sem[2].signal, proc.sec}, | |
| | {A.proc.ok_sig, proc.sec} | -> {proc.stop}, | |
| **end;** | | | |
| **server:** | r (**agents**:A3), | | |
| **services** | {left, right}, | | |
| **states** | {res}, | | |
| **actions** | {A3.r.left, r.res} | -> {A3.r.right, r.res}, | |
| | {A3.r.right, r.res} | -> {A3.r.left, r.res}, | |
| **end;** | | | |

| | |
|---|---|
| **agents:** | A[2], A3; |
| **servers:** | sem[2], proc[2], r; |
| **init** -> | { |
| | <j=1..2>A[j].proc[j].start, |
| | A3.r.left, |
| | <j=1..2>proc[j](A[j],sem[j],sem[3-j]).ini,; |
| | <j=1..2>sem[j](A[1],A[2],proc[1],proc[2]).up, |
| | r(A3).res, |
| | }. |

Note that the system above is defined in the resident view (server view), as the actions concerning a given server are grouped. Server states: *up, down, firs, sec* are hidden inside resident processes (not used outside the containing servers), and the communication is performed via messages (*wait, ok_wait, …*). The system may be automatically converted to the agent view (because the decomposition is canonical). In the traveler view (agent view) the messages are hidden inside travelers, they communicate via shared server states.

| | | | |
|---|---|---|---|
| **agent**: | A (**servers**:proc,sem[2]), | | |
| **services** | {start, ok_wait, ok_sig, wait, signal}, | | |
| **states** | {ini, first, sec, stop, up, down}, | | |
| **actions** | {A.proc.start, proc.ini} | -> {A.sem[1].wait, proc.first}, | |
| | {A.sem[1].wait, sem[1].up} | -> {A.proc.ok_wait, sem[1].down}, | |
| | {A.proc.ok_wait, proc.first} | -> {A.sem[2].wait, proc.sec}, | |
| | {A.sem[2].wait, sem[2].up} | -> {A.proc.ok_wait, sem[2].down}, | |
| | {A.proc.ok_wait, proc.sec} | -> {A.sem[1].signal, proc.first}, | |
| | {A.sem[1].signal, sem[1].down} | -> {A.proc.ok_sig, sem[1].up}, | |

|                    | {A.proc.ok_sig, proc.first}      | -> {A.sem[2].signal, proc.sec},     |
|                    | {A.sem[2].signal, sem[2].down}   | -> {A.proc.ok_sig, sem[2].up},      |
|                    | {A.proc.ok_sig, proc.sec}        | -> {proc.stop},                     |

**end**;

**agent**: A3 (**servers**:r),
**services** {left, right},
**states** {res},
**actions** {A3.r.left, r.res}    -> {A3.r.right, r.res},
{A3.r.right, r.res}   -> {A3.r.left, r.res},
**end**;

**agents**: A[2], A3;
**servers**: sem[2], proc[2], r;

**init** -> {
    <j=1..2>proc[j].ini,
    <j=1..2>sem[j].up,
    r.res,

    <j=1..2>A[j](proc[j],sem[j],sem[3-j]).start,
    A3(r).left,
}.

The LTL formula describing that the process *A[1]* falls into deadlock may be expressed as
$\Diamond (P_{A[1]} \land \Box \sim A_{A[1]})$ where $P_{A[1]}$ is a set of all system states in LTS (recall that a system state is a maximal system configuration) containing messages with agent identifier *A[1]* (in general, agents may be created and annihilated), and $A_{A[1]}$ is a set of all actions with these messages on input. The formula reads "there is possible a situation in which $P_{A[1]}$ is a current system state in which a message with *A[1]* agent identifier is pending, and no action belonging to $A_{A[1]}$ may be executed in the future". This formula is true of course (a deadlock in the traveler *A[1]* may occur in the system). In a case of absence of *A3* it would be a total deadlock (when *A[1]* is in deadlock, *A[2]* falls into a deadlock as well). But in our system (where *A3* is present) the deadlock is partial. Yet the formula and its verification will remain the same in both systems: with *A3* and without *A3*. The important feature of this scheme of verification is that it does not use any particular item, it says: "the process *A[1]* falls into a deadlock", whenever it occurs. A counterexample generated by model checker identifies the point of deadlock.

Promela code generated for the server view of the example system is:

```
mtype = {none, start, ok_wait, ok_sig, wait, signal, left, right};

//sem_serv = {start, ok_wait, ok_sig}; -> mtype
//sem_val = {up, down};
#define up 1
#define down 2
mtype A1serv[3], A2serv[3];
int semact[3];
#define sem1cA1 A1serv[1]!=none
#define sem1cA2 A2serv[1]!=none
#define sem2cA1 A1serv[2]!=none
#define sem2cA2 A2serv[2]!=none
proctype sem (int num; int inival; chan A1, A2, A1_proc1, A2_proc2)
{       mtype mes;
        int state;
        state=inival;
        //mtype A1serv, A2serv;
```

```
            A1serv[num]=none;
            A2serv[num]=none;
            semact[num]=0;
            do
            :: A1?<mes> -> A1serv[num]=mes;
              if
              :: (mes==wait) && (state==up) ->
                    A1?mes; semact[num]=1; state=down;
                    A1serv[num]=none; semact[num]=0; A1_proc1!ok_wait
              :: (mes==signal) && (state==down) ->
                    A1?mes; semact[num]=1; state=up;
                    A1serv[num]=none; semact[num]=0; A1_proc1!ok_sig
              :: (mes==signal) && (state==up) ->
                    A1?mes; semact[num]=1; state=up;
                    A1serv[num]=none; semact[num]=0; A1_proc1!ok_sig
              fi
            :: A2?<mes> -> A2serv[num]=mes;
              if
              :: (mes==wait) && (state==up) ->
                    A2?mes; semact[num]=1; state=down;
                    A2serv[num]=none; semact[num]=0; A2_proc2!ok_wait
              :: (mes==signal) && (state==down) ->
                    A2?mes; semact[num]=1; state=up;
                    A2serv[num]=none; semact[num]=0; A2_proc2!ok_sig
              :: (mes==signal) && (state==up) ->
                    A2?mes; semact[num]=1; state=up;
                    A2serv[num]=none; semact[num]=0; A2_proc2!ok_sig
              fi
            od
}

//proc_serv = {wait, signal}; -> mtype
#define ini 1
#define first 2
#define sec 3
#define stop 4
//proc_val = {ini, first, sec, stop};
mtype Aserv[3];
#define proc1cA Aserv[1]>none
#define proc2cA Aserv[2]>none
proctype proc (int num; int inival; chan A, A_sem1, A_sem2)
{       mtype mes;
        int state;
        state=inival;
        //mtype Aserv;
        Aserv[num]=none;
        Aserv[num]=none;
        do
        :: A?<mes> -> Aserv[num]=mes;
          if
          :: (mes==start) && (state==ini) ->
                A?mes; procact[num]=1; state=first;
                Aserv[num]=none; procact[num]=0;  A_sem1!wait
          :: (mes==ok_wait) && (state==first) ->
                A?mes; procact[num]=1; state=sec;
                Aserv[num]=none; procact[num]=0;  A_sem2!wait
          :: (mes==ok_wait) && (state==sec) ->
                A?mes; procact[num]=1; state=first;
                Aserv[num]=none; procact[num]=0;  A_sem1!signal
          :: (mes==ok_sig) && (state==first) ->
```

```
                        A?mes; procact[num]=1; state=sec;
                        Aserv[num]=none; procact[num]=0;  A_sem2!signal
                :: (mes==ok_sig) && (state==sec) ->
                        A?mes; procact[num]=1; state=stop;
                        Aserv[num]=none; procact[num]=0
                fi
        od
}

//r_serv = {left, right}; -> mtype
#define res 1
//r_val = {res};
mtype A3serv;
#define rcA3 A3serv!=none
proctype r (int inival; chan A3)
{       mtype mes;
        int state;
        state=inival;
        //mtype A3serv;
        A3serv=none;
        do
        :: A3?<mes> -> A3serv=mes;
          if
          :: (mes==left) && (state==res) ->
                  A3?mes; ract[num]=0;  state=res;
                   A3serv=none; ract[num]=1;  A3!right
          :: (mes==right) && (state==res) ->
                  A3?mes; ract[num]=0;  state=res;
                  A3serv=none; ract[num]=1;  A3!left
          fi
        od
}

init {
        chan A1_sem1 = [1] of {mtype};
        chan A1_sem2 = [1] of {mtype};
        chan A2_sem1 = [1] of {mtype};
        chan A2_sem2 = [1] of {mtype};
        chan A1_proc1 = [1] of {mtype};
        chan A2_proc2 = [1] of {mtype};
        chan A3_r = [1] of {mtype};

        run proc(1,ini,A1_proc1,A1_sem1,A1_sem2);
        run proc(2,ini,A2_proc2,A2_sem2,A2_sem1);
        run sem(1,up,A1_sem1,A2_sem1,A1_proc1,A2_proc2);
        run sem(2,up,A1_sem2,A2_sem2,A1_proc1,A2_proc2);
        run r(res,A3_r);

        A1_proc1!start;
        A2_proc2!start;
        A3_r!left;

}
```

# 6   Two Kinds of Deadlock and Distributed Termination

Deadlock over resources is defined in IMDS as discontinuation of a traveler process with given agent identifier (but not in terminating action). In our approach, we express deadlock over resources as general formula:

For an $a \in AGT$, the traveler with agent identifier $a$ is in deadlock over resources iff
$$\Diamond (M_a \wedge \Box \sim A_a), \tag{33}$$
where:
- $M_a$ is a set of all messages of agent $a$,
- $A_a$ is a set of all actions with these messages on input.

This was described in a previous chapter on the example of agent $A[1]$ as $a$.

The system may be dually described by resident processes, formed from actions following server states. In this view, deadlock over communication may be described[4]. A communication deadlock means that no action on a server may be executed while there are messages pending, having this server identifier. The general formula for communication deadlock is:

For a $svr \in SVR$, the resident with identifier $svr$ is in communication deadlock iff
$$\Diamond (M_{svr} \wedge \Box \sim A_{svr}), \tag{34}$$
where:
- $M_{svr}$ is a set of all messages with server identifier $svr$,
- $A_{svr}$ is a set of all actions with states of server $svr$ on input.

It is important that in IMDS we can differentiate deadlock from distributed termination [Chro00]. For static traveler process (whose messages are present in initial configuration), termination is simply a situation in which the agent disappears (no continuation message is generated in an action belonging to the agent). To be sure that the termination is inevitable, the operator $\Diamond$ (eventually) should be used for a set $M_a$ of all messages of the agent $a$: $\Diamond \Box \sim M_a$ (eventually always not $M_a$). For dynamic process (whose message is generated is some action) the formula should read that once an agent $a$ occurs, it must disappear (this works for static process as well). Therefore, the general formula for traveler termination is:

For an $a \in AGT$, the traveler with agent identifier $a$ terminates iff
$$\Box (M_a \Rightarrow \Diamond \Box \sim M_a), \tag{35}$$
where $M_a$ is a set of all messages with agent identifier $a$.

All proposed formulas, as expressed in LTL, require only linear time (to the size of a state space) to be evaluated. Moreover, the formulas do not change independently of presence or absence of other processes in a system (only a given agent $a$ or a given server $svr$ is used in formulas).

In our example, tests for deadlock and termination give the following results:

$\Diamond (M_{A[1]} \wedge \Box \sim A_{A[1]})$ (traveler $A[1]$) – true (deadlock over resources in $A[1]$)
$\Diamond (M_{A[2]} \wedge \Box \sim A_{A[2]})$ (traveler $A[2]$) – true (deadlock over resources in $A[2]$)
$\Diamond (M_{A3} \wedge \Box \sim A_{A3})$ (traveler $A3$) – false (no deadlock over resources in $A3$)

$\Diamond (M_{sem[1]} \wedge \Box \sim A_{sem[1]})$ (resident $sem[1]$) – true (communication deadlock in $sem[1]$)
$\Diamond (M_{sem[2]} \wedge \Box \sim A_{sem[2]})$ (resident $sem[2]$) – true (communication deadlock in $sem[2]$)
$\Diamond (M_r \wedge \Box \sim A_r)$ (resident $r$) – false (no communication deadlock in $r$ – actions executing *left* and *right* services identifier are executed forever)

---

[4] Not every deadlock over resources is a communication deadlock, as on a server some sequence of items with the same agent identifier may stop, while other sequences are alive. In [Chro01], a classification of deadlocks is presented.

Also, communication deadlock does not happen to *proc[1]* and *proc[2]* servers – they simply wait for next action to be executed:

$\lozenge (M_{proc[1]} \land \square \sim A_{proc[1]})$ (resident *proc[1]*) – false
$\lozenge (M_{proc[2]} \land \square \sim A_{proc[2]})$ (resident *proc[2]*) – false

$\square (P_{A[1]} \Rightarrow \lozenge \square \sim P_{A[1]})$ – false (traveler *A[1]* may not terminate)
$\square (P_{A[2]} \Rightarrow \lozenge \square \sim P_{A[2]})$ – false (traveler *A[2]* may not terminate)
$\square (P_{A3} \Rightarrow \lozenge \square \sim P_{A3})$ – false (traveler *A3* does not terminate, which is desired as it contains an endless loop)

To sum up, in our case traveler processes with agent identifiers *A[1]* and *A[2]* fall into deadlock over resources while traveler *A3* does not. Also, resident processes on servers *sem[1]* and *sem[2]* fall into communication deadlock while *r* does not. No traveler necessarily terminates.

# 7  Conclusions

The small repertoire of behavior features in distributed systems that may be verified by model checking (especially connected with deadlock and termination). It comes from the lack of formalisms in which some detailed local features may be expressed. The IMDS formalism for description of distributed systems liberates the designer from the simple "state with no future" verification which allows to find only global deadlock/termination states and gives no help how to distinguish them from each other.

The application of IMDS formalism allows to highlight locality of features and differentiates between features relative to travelers behavior (in agent view) from features relative to residents (in server view). IMDS (where no global system state is observed in execution of an action) is especially destined to true distributed systems, like LANs, internet services, multiagent systems. The locality of "observations" allows to verify a part of a system (a subset of servers or a subset of agents) independently on the activities of the rest of the system, possibly changing in its various versions. Therefore, real negative and positive features of distributed systems may be verified using the proposed idea.

To check communication deadlocks, resource deadlocks (including partial deadlocks) and termination in an arbitrary system, using the methodology presented in the paper, the designer should:
- express the system (or – better – its model; or best – the communication skeleton of the system) in terms of IMDS actions (of course servers, valuations, services and agents should be defines first); the proposed notation is presented informally in the paper – it is intuitional form – simply input and output items of any actions are listed;
- encode the set of IMDS actions in input notation of a model checker;
- apply temporal formulas, for example: $\lozenge (M_a \land \square \sim A_a)$ for every agent *a* (deadlocks over resources), $\lozenge (M_{svr} \land \square \sim A_{svr})$ for every server *ser* (communication deadlocks), $\square (M_a \Rightarrow \lozenge \square \sim M_a)$ for every agent *a* (termination)[5].

The combination of IMDS with model checking allows to express and verify variety of features like deadlock over resources, communication deadlock, partial deadlock, termination

---
[5] for computation tree logic CTL the formulas should have the form: **EF** $(P_a \land$ **AG** $\sim A_a)$, **EF** $(P_{ser} \land$ **AG** $\sim A_{ser})$, **AG** $(P_a \Rightarrow$ **AF AG** $\neg P_a)$

etc. There are many more classes of distributed behavior observed than in "classical" model checking approach. Note that all proposed formulas are evaluated in time linear to the size of the model.